\documentclass[aps,prd,preprint,nofootinbib,showpacs,superscriptaddress]{revtex4}
\usepackage{epsfig}
\usepackage{graphicx}
\usepackage{dsfont}
\usepackage{amsmath}

\newcommand{\eg}{\emph{e.g.}}
\newcommand{\ie}{\emph{i.e.}}
\newcommand{\Ref}{Ref.}
\newcommand{\Refs}{Refs.}
\newcommand{\fig}{Fig.}

\newcommand{\eq}{Eq.}
\newcommand{\bea}{\begin{eqnarray}}
\newcommand{\eea}{\end{eqnarray}}
\newcommand{\nn}{\nonumber}

\newcommand{\Sec}{Sec.}
\newcommand{\eps}{\varepsilon}
\newcommand{\CP}{\emph{CP}}
\newcommand{\mue}{\mu e}
\newcommand{\sig}[1]{#1\sigma}

\DeclareMathOperator{\tr}{Tr}

\begin{document}
\title{Non-standard interaction effects on astrophysical neutrino fluxes}

\author{Mattias Blennow}
\email[]{blennow@mppmu.mpg.de}
\affiliation{Max-Planck-Institut f\"ur Physik
(Werner-Heisenberg-Institut), F\"ohringer Ring 6, 80805 M\"unchen,
Germany}

\author{Davide Meloni}
\email[]{meloni@fis.uniroma3.it}
\affiliation{Dipartimento di Fisica, Universit\'a di Roma Tre and INFN Sez.~di Roma Tre,
via della Vasca Navale 84, 00146 Roma, Italy}

\date{\today}

\preprint{MPP-2009-4}
\preprint{RM3-TH/09-2}

\begin{abstract}
We investigate new physics effects in the production and detection of
high energy neutrinos at neutrino telescopes.  Analysing the flavor ratios 
$\phi_\mu/\phi_\tau$ and $\phi_\mu/(\phi_\tau+\phi_e)$, we find that the
Standard Model predictions for them can be sensibly altered by new
physics effects.

\end{abstract}
\pacs{14.60.Lm, 14.60.Pq, 95.55.Vj, 95.85.Ry}
\maketitle

\section{Introduction}
\label{sec:intro}

Neutrino oscillation physics has definitively entered the era of
precision measurements of the fundamental neutrino parameters. Recent
experiments such as Super-Kamiokande, SNO, KamLAND, K2K, and MINOS
\cite{Hosaka:2006zd,Ahmed:2003kj,Oblath:2007zz,Ahn:2006zza,Michael:2006rx,:2008ee},
have improved our knowledge of the neutrino mass squared differences
(\ie, $\Delta m_{31}^2$ and $\Delta m_{21}^2$) and some of the the
leptonic mixing parameters (\ie, $\theta_{12}$, $\theta_{23}$). Whether
$\theta_{13}$ is different from zero and \CP-violation is present in
the leptonic sector of the Standard Model are questions which will be
addressed by forthcoming experiments \cite{Bandyopadhyay:2007kx}.

While there exist clear evidence that the $V-A$ structure of the weak
interactions of the Standard Model correctly describes neutrino
interactions with matter, there is still a possibility that some
next-to-leading order mechanism affects the processes of neutrino
production and detection. In general, this sort of Physics beyond the
Standard Model is described by a set of higher-dimensional
non-renormalizable operators suppressed by some high energy scale
(see, \eg, \Refs~\cite{Antusch:2008tz,Gavela:2008ra} and references
therein). The precision measurements in the neutrino sector then open
up the possibility to investigate such non-standard interactions (NSI)
at a quite accurate level.
  
Neutrino oscillations and NSI in terrestrial neutrino experiments have
been studied extensively in the literature, using the neutrino factory
project \cite{NSInfstart,Ota:2001pw,GonzalezGarcia:2001mp,Gago:2001xg,
  Huber:2001zw,Kopp:2007mi,Ribeiro:2007ud,Kopp:2008ds,Winter:2008eg,Altarelli:2008yr,NSInfstop}
and other different neutrino facilities (\eg, conventional neutrino beams, super-beams and
$\beta$-beams)
\cite{NSIotherstart,Adhikari:2006uj,Blennow:2007pu,Kopp:2007ne,Blennow:2008ym,Ohlsson:2008gx,NSIotherstop}
to assess the impact of the NSI in neutrino physics.

Here we adopt a different point of view, we investigate the NSI in the
neutrino sector using very high energy neutrinos from astrophysical
sources and the capability of neutrino telescopes to measure their
fluxes on Earth \cite{icecube,baikal,antares,nestor,nemo}. We rely on the simplified assumption that the new
physics effects arise in the production and detection processes but do
not affect the neutrino propagation. We also assume that the lepton
mixing matrix is the standard unitary matrix describing the couplings
of the charged-lepton--neutrino--$W$ vertices. Moreover, for the sake
of illustration, we prefer to study the possible NSI signals for the
three different sources (pion, muon damped and neutron sources)
separately, as the effects of new physics are quite different in these
cases.

This work is organized as follows. In \Sec~\ref{sec:SDNSI}, we will
present analytic considerations for the NSI we study in the paper. In
particular, we will first describe the detector effects in a unified
way, since these are independent of the assumed neutrino source. We
will then address the question of non-standard physics in the
production processes.  In \Sec~\ref{sec:exp}, we will describe the
statistical approach we use to study the sensitivity of neutrino
telescopes to new physics effects and also present our numerical
results. Finally, in \Sec~\ref{sec:S&C}, we will present a summary of
the work as well as our conclusions.

\section{NSI at the source and detector}
\label{sec:SDNSI}

We consider NSI through effective four-fermion operators of the form
\begin{equation}
  \mathcal L_{\rm B-NSI} = - 2\sqrt 2 G_F \cos(\theta_W) \eps^{ud}_{\alpha\beta} [\bar u \gamma^\mu P_L d][\bar \ell_\alpha \gamma_\mu P_L \nu_\beta] + {\rm h.c.}
\end{equation}
for charged-current NSI with baryons, where $\theta_W$ is the Weinberg
angle, and
\begin{equation}
  \mathcal L_{\ell-\rm NSI} = -2\sqrt 2 G_F \eps^{\alpha\beta}_{\gamma\delta}
    [\bar \ell_\alpha \gamma^\mu P_L \ell_\beta][\bar \nu_\gamma \gamma_\mu P_L \nu_\delta]
\end{equation}
for purely leptonic NSI (see \cite{Barranco:2007ej,Davidson:2003ha,Berezhiani:2001rs,Barranco:2005ps} for a discussion on the limits on $\eps^{\alpha\beta}_{\gamma\delta}$). While the addition of the hermitian conjugate
in the case of NSI with baryons leads to operators of the form $[\bar
  d \gamma^\mu P_L u][\bar \nu_\beta \gamma_\mu P_L \ell_\alpha]$, the
requirement of a hermitian Lagrangian for the purely leptonic NSI
implies $\eps^{\alpha\beta}_{\gamma\delta} =
\eps^{\beta\alpha*}_{\delta\gamma}$. In what follows, we will assume
that the coherence among the neutrino mass eigenstates that arrive at
the Earth has been lost. Thus, we do not need to take any interference
terms into account and may treat the source and detector processes
separately.

\subsection{Detector effects}

Since the detector processes are essentially the same regardless of
the source, we will start by considering these. In principle, the
detection of astrophysical neutrinos is done through charged-current
reactions of the form $\nu + X \rightarrow Y + \ell_\alpha$ and then
identifying the neutrino flavor through identification of the outgoing
charged lepton. Assuming that the neutrino arrives at the Earth in the
mass eigenstate $\nu_i$, the matrix element involved in computing the
reaction rate is given by
\begin{equation}
  \mathcal M_{i\alpha} = \mathcal M_0 [(\mathds 1 + \eps^{ud})U]_{\alpha i},
\end{equation}
where $\mathcal M_0$ is the matrix element for the corresponding
reaction if the incoming neutrino would have been of flavor $\alpha$
and there were no NSI, $\mathds 1$ is the $3 \times 3$ unit matrix,
$\eps^{ud} = (\eps^{ud}_{\alpha\beta})$, and $U$ is the leptonic
mixing matrix. Thus, if the incoming $\nu_i$ flux is $\phi_i$, then
the measured flux of $\nu_\alpha$ is given by
\begin{equation}
  \phi_\alpha = \phi_i |[(\mathds 1 + \eps^{ud})U]_{\alpha i}|^2.
\end{equation}

Naturally, the composition of the neutrino flux arriving at the Earth
is not purely $\nu_i$. Thus, in order to compute the
actual flavor flux, we need to sum over all mass eigenstates and
arrive at
\begin{equation}
  \label{eq:detector}
  \phi_\alpha = \sum_i \phi_i |[(\mathds 1 + \eps^{ud})U]_{\alpha i}|^2.
\end{equation}
The actual fluxes $\phi_i$ are dependent on the source type and may
also contain effects from the same NSI as those affecting the detector
processes.

\subsection{Hadronic sources}

Let us first assume that the astrophysical neutrino source creates
neutrinos through hadron decays only. Out of the usually considered
scenarios, this category includes the muon-damped pion sources (giving
initial flavor ratios of $\phi_e:\phi_\mu:\phi_\tau = 0:1:0$ in the
neutrino fluxes) as well as neutron-like sources ($1:0:0$). These
sources produce neutrinos through processes where a meson or baryon decays into
a charged lepton of a given flavor\footnote{In the case of pion decay,
  there is a small contamination of decays into electrons which is
  suppressed by $m_e^2/m_\mu^2$.}, a neutrino and possibly other
products. When a neutrino is produced together with a charged lepton
of flavor $\beta$ in a weak decay of a meson or a baryon, the probability of
producing it in the neutrino mass eigenstate $\nu_i$ is given by
\begin{equation}
  P_i = |U_{\beta i}|^2
\end{equation}
in the Standard Model. With the addition of NSI, there
is also the possibility to produce the neutrino in a different flavor
state and the corresponding probability changes according to
\begin{equation}
  P_i \propto |[(\mathds{1}+\eps^{ud})U]_{\beta i}|^2.
\end{equation}
In general, the full expression for $P_i$ also contains a
normalization factor, since $\mathds 1 + \eps^{ud}$ may not be
unitary. However, this normalization factor is the same for all $P_i$
and will be removed once we consider neutrino flux ratios. Since the
meson/baryon decay gives the only neutrino contribution in this scenario,
the $\nu_i$ flux will be given by
\begin{equation}
  \phi_i = \phi_0 P_i,
\end{equation}
where $\phi_0$ is the total initial neutrino flux. By insertion into
\eq~(\ref{eq:detector}), the measured flux of $\nu_\alpha$ at the
detector is therefore given by
\begin{equation}
\label{eqgen}
\phi_\alpha \propto \phi_0 \sum_i \bigg|[(\mathds{1}+\eps^{ud})U]_{\beta i}\bigg|^2 
\bigg|[(\mathds{1}+\eps^{ud})U]_{\alpha i}\bigg|^2.
\end{equation}

\subsection{Non muon-damped pion sources}

We now consider the situation where the source is producing neutrinos
both through an initial decay of a charged pion and through the
subsequent decay of the resulting charged muon. The complete decay
chain is then
\begin{equation}
\begin{array}{rcl}
\pi^+ \longrightarrow & \mu^+& + \nu_\alpha \\
                      & \downarrow & \\
                      & e^+ & + \nu_\beta + \bar\nu_\gamma
\end{array},
\end{equation}
as well as the corresponding \CP-conjugate reaction for $\pi^-$. With
only Standard Model interactions, $\alpha = \gamma = \mu$ and $\beta =
e$; the produced flavor ratio is then approximately $1:2:0$.

Mathematically, the decay of the pion can be described in the same way
as in the previous section, with the exception that we now need to
know the normalization factor for $P_i$, since we will add the
contributions from the pion and muon decays and need to be consistent
while doing so. The normalization factor $N_\mu$ is simply given by
the fact that the total probability of the pion decay should be equal
to one, \ie,
\begin{equation}
  \sum_i P_i = \frac 1{N_\mu} \sum_i |[(\mathds{1}+\eps^{ud})U]_{\mu i}|^2 = 1
  \quad \Longrightarrow \quad
  N_\mu = [(\mathds 1 + \eps^{ud})(\mathds 1 + \eps^{ud\dagger})]_{\mu\mu}
\end{equation}
(essentially, this is the factor by which the decay rate of the pion
would change due to the NSI).

The muon decay is a little more involved, since the final state
involves two neutrinos, where one of these is not observed. In the
literature, this problem is usually solved by considering only NSI of
the form $[\bar e \gamma^\rho P_L \mu][\bar \nu_\alpha \gamma_\rho P_L
  \nu_e]$. With this simplification, the situation is completely
analogous to the case of the pion decay, simply because we know the
flavor of the outgoing $\bar\nu$ (in the case of $\mu^+$-decay). With
general NSI however, we need to consider the full matrix element for
the decay $\mu^+ \to e^+\bar\nu_i \nu_j$, which is given by
\begin{equation}
  \mathcal M_{ij} = \mathcal M_0 [U^\dagger (\mathcal J^{\mu e} + \eps^{\mu e}) U]_{ji}
\end{equation}
where $\mathcal M_0$ is the matrix element when neutrinos do not mix
and only Standard Model interactions are considered, $\mathcal
J^{\mue}_{\alpha\beta} = \delta_{e\alpha}\delta_{\mu\beta}$, and
$\eps^{\mu e} = (\eps^{\mu e}_{\alpha\beta})$ is a matrix containing
the strengths of the NSI. In order to obtain the probability for a
neutrino from this type of decay to be in the mass eigenstate $\nu_i$,
we need to consider the fact that we do not measure the anti-neutrino
from the same decay. Thus, the probability will be given by an
incoherent sum over the antineutrino mass eigenstates as
\begin{equation}
  P^{\mu^+}_i \propto \sum_j |\mathcal M_{ij}|^2 \propto 
  [U^\dagger (\mathcal J^{\mue}+\eps^{\mue})^\dagger (\mathcal J^{\mue}+\eps^{\mue})U]_{ii}.
\end{equation}
Again, this needs to be normalized and the normalization factor is given by
\begin{equation}
  N^{\mu^+} = 
  \tr[ (\mathcal J^{\mue}+\eps^{\mue})^\dagger (\mathcal J^{\mue}+\eps^{\mue})]
\end{equation}
The corresponding argumentation for the outgoing antineutrino results in
\begin{equation}
  \bar P^{\mu^+}_{j} \propto \sum_i |\mathcal M_{ij}|^2 \propto 
  [U^\dagger (\mathcal J^{\mue}+\eps^{\mue}) (\mathcal J^{\mue}+\eps^{\mue})^\dagger U]_{jj}
\end{equation}
with the same normalization constant $\bar N^{\mu^+} =
N^{\mu^+}$. Repeating the same derivation for $\mu^-$-decay, we arrive
at $P^{\mu^-}_i = \bar P^{\mu^+}_i$ and $\bar P^{\mu^-}_j =
P^{\mu^+}_j$. Thus, assuming equal numbers of positive and negative
pions decaying, the flux of neutrino mass eigenstate $\nu_i$ (or
antineutrino mass eigenstate $\bar \nu_i$) is given by
\begin{equation}
\phi_i = \phi_0 (P_i + P^{\mu^+}_i + \bar P^{\mu^+}_i),
\end{equation}
where $\phi_0$ is the initial flux of $\nu_e$ in the case of Standard Model
interactions only.

\section{Experimental implications}
\label{sec:exp}

For experimental reasons, it is customary to consider the flavor flux
ratios
\begin{equation}
\label{ratios}
R_{e\mu} = \frac{\phi_e}{\phi_\mu}, \quad
R_{\mu\tau} = \frac{\phi_\mu}{\phi_\tau}, \quad {\rm and} \quad
R = \frac{\phi_\mu}{\phi_e + \phi_\tau}
\end{equation}
(note that these are not independent). With the flavor fluxes computed
as in \Sec~\ref{sec:SDNSI}, we want to know how NSI at the source and
detector can affect the results of these measurements. In order to do
this, we will examine how well a measured value of any of the $R$s can
be accommodated when considering Standard Model interactions only, as
well as when allowing for NSI. To quantify this, we introduce the
$\chi^2$ function
\begin{equation}
\label{chi2def}
  \chi^2(x) = \sum_i \left(\frac{R^{\rm exp}_i-R^{\rm th}_i(x)}{\sigma_{R_i}}\right)^2
     + \sum_j \left(\frac{x_j - x_{j,0}}{\sigma_{x_j}}\right)^2.
\end{equation}
Here, $R^{\rm exp}_i$ is the measured value of a given ratio, $R^{\rm
  th}_i(x)$ is the theoretically expected value computed from the set
$x$ of input parameters, $\sigma_{R_i}$ is the experimental
uncertainty in $R_i$, $i$ is an index running over the flux ratios
considered, $\sigma_{x_j}$ are the external uncertainties for the input
parameters, and $j$ is an index running over the input parameters. In
the case of Standard Model interactions only, $x$ contains the
neutrino mixing parameters, while in the NSI scenario it also contains
all of the different $\eps$s. How well a particular measured
combination of $R^{\rm exp}$s can be accommodated within a model is then
quantified by
\begin{equation}
  \chi^2_{\rm min} = \min_{x}[\chi^2(x)],
\end{equation}
which is $\chi^2$ distributed with $n$ degrees of freedom, where $n$
is the number of independent flavor flux ratios considered.

In the following numerical computations, we adopt the standard
parametrization of the neutrino mixing matrix (see
\Refs~\cite{Pontecorvo:1957cp,Maki:1962mu,Pontecorvo:1967fh,Gribov:1968kq,Chau:1984fp}),
assuming also a tri-bimaximal structure \cite{Harrison:2002er} for it.
We assume the $\sig{3}$ errors on the standard mixing parameters to be
of the same order as the errors in \Ref~\cite{Maltoni:2004ei}. In
Tab.~\ref{tab:nufit}, we give the corresponding values for these
parameters.
\begin{table} 
\begin{center}
\begin{tabular}{||c|c|c||}
\hline
\hline
 {\it Parameter} & {\it Best fit} & {\it $\sig{3}$ error} \\ 
\hline
$\theta_{12}$ & $35.3^\circ$ & $5^\circ$ \\ 
\hline
$\theta_{13}$  & 0 & $12.5^\circ$ \\
\hline
$\theta_{23}$& $45^\circ$  & $10^\circ$ \\
\hline
$\delta$ & \multicolumn{2}{|c||}{Free ($[0,2\pi]$)} \\
\hline
\hline
\end{tabular}
\end{center}
\caption{\it \label{tab:nufit} Summary of the best fit values and
  $\sig{3}$ errors used in our numerical computation for the standard
  neutrino oscillation parameters.}
\end{table}

\subsection{Muon-damped pion sources}
\begin{figure}
\begin{center}
\includegraphics[width=0.75\textwidth,clip=true]{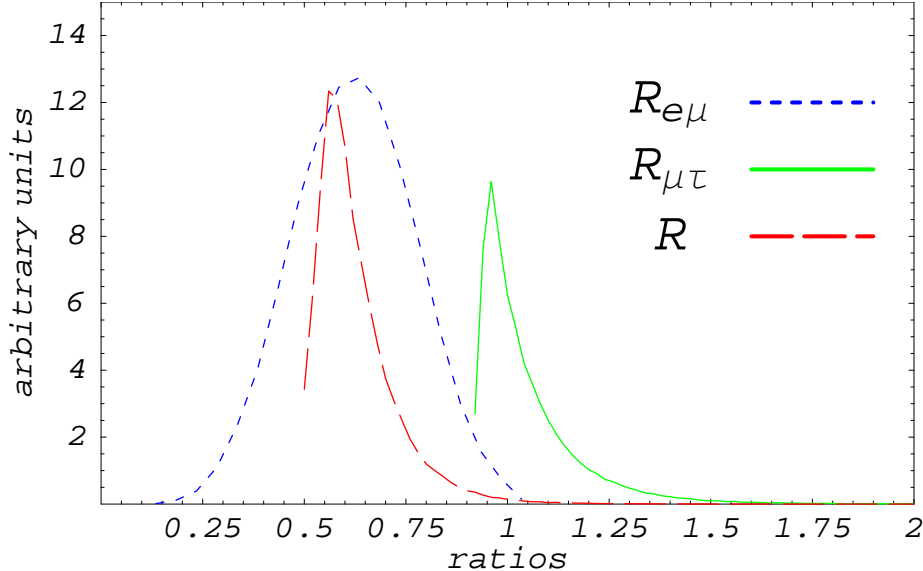}
\caption{\it Statistical distribution (in arbitrary units) of the flux
  ratios $R_{e\mu}$ (dashed line), $R_{\mu\tau}$ (solid line) and $R$
  (long-dashed line), as computed in the Standard Model for a
  muon-damped pion source.}
\label{fig:distr}
\end{center}
\end{figure}

In order to appreciate the effects of the new physics on the flux
ratios of \eq~(\ref{ratios}), it is useful to recall the Standard
Model expectations for these quantities.  Putting $\eps^{ud}=0$ and
$\beta=\mu$ in \eq~(\ref{eqgen}), we get the following expressions for
the flavor fluxes: 
\bea \phi_e 
 &=&
 \phi_0\,\sum_i |U_{\mu i}|^2 |U_{ei}|^2, \nn \\ 
\phi_\mu
 &=&
 \phi_0\,\sum_i |U_{\mu i}|^4, \nn \\ 
\phi_\tau
 &=&
 \phi_0\,\sum_i |U_{\mu i}|^2 |U_{\tau i}|^2 .
\eea
In the limit of exact tri-bimaximal mixing ($\theta_{13}=0$, $\sin^2
\theta_{12}=1/{3}$, and $\sin^2 \theta_{23}=1/{2}$
\cite{Harrison:2002er}), we obtain:
\begin{equation}
R_{e\mu} = \frac{4}{7}, \quad
R_{\mu\tau} = 1, \quad {\rm and} \quad
R = \frac{7}{11}.
\end{equation}
Before concluding that large deviations from these values are signals
of new physics, we should carefully take the role of the uncertainties
of the standard parameters into account \cite{Meloni:2006gv}. In order
to do that, we compute the distributions of the three ratios,
extracting neutrino mixing angles according to Gaussian distributions
with central values and $\sig{1}$ error as deducible from
Tab.~\ref{tab:nufit}. The result of this procedure is shown in
\fig~\ref{fig:distr}.
It can be clearly seen that the larger spread is obtained for
$R_{e\mu}$, due to the uncertainties on the mixing angles
$\theta_{12}$, $\theta_{23}$ and on the product $\cos \delta
\,\theta_{13}$ (see also \eq~(4) in \Ref~\cite{Lipari:2007su}), which
are all of the same order of magnitude. On the other hand, the fact
that most decays will result in neutrinos of the mass eigenstate with
the largest $\nu_\mu$ content protects $R_{\mu\tau}$ from becoming
much smaller than one, an effect that can also be seen in the ratio
$R$, which cannot be much smaller than $1/2$.  The pile-up of the
distributions close to these values are the results of $\theta_{23}$
being close to maximal. It is then possible that new physics effects
would be much more visible in the ratios containing $\tau$ neutrinos
and, for this reason, we will investigate the effects of the $\eps$
parameters in the $R_{\mu\tau}$--$R$-plane.

\begin{figure}
\begin{center}
\includegraphics[width=0.75\textwidth,clip=true]{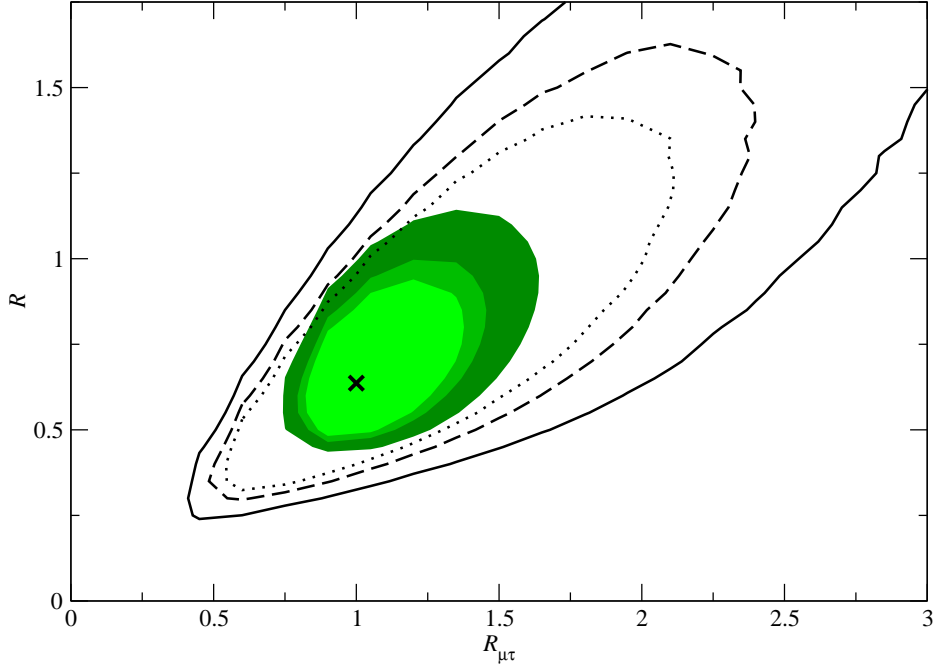}
\caption{\it Isocontours of $\chi^2_{\rm min}$ in the
  $R_{\mu\tau}$--$R$-plane in the case of a muon damped pion source in
  the standard framework for neutrino oscillation and interaction
  (shaded regions) as well as for the framework with included NSI
  (black curves). The contours correspond to 90~\%, 95~\%, and 99~\%
  C.L., respectively. The black cross corresponds to the prediction in
  the case of tribimaximal mixing in the standard framework.}
\label{fig:MDFULL}
\end{center}
\end{figure}

The correlation between $R_{\mu\tau}$ and $R$ can be analyzed using
the $\chi^2$ definition in \eq~(\ref{chi2def}). A crucial point here
is the precision we expect in the experimental measurement of the flux
ratios, $\sigma_{R_i}$; for the sake of illustration, we will assume
that $\sigma_{R_i}=0.1 R_i^{\rm exp}$ for any ratio (the effect of
changing this value is discussed in \Sec~\ref{sec:unc}), which means
that they are measured with a 10~\% error. In addition, we assume an
external error of 0.1 for the $\eps$ parameters. The reason for
choosing such a large error is that we are mainly interested in the
qualitative impact of the parameters rather than making precise
predictions. The results of the minimization of the $\chi^2$ over all
model parameters is shown in \fig~\ref{fig:MDFULL}.
We note that the extension of the isocontours produced by new physics
is predominantly in the direction of large ratios; in particular,
$R_{\mu\tau}$ and $R$ can be as large as twice their tri-bimaximal
mixing values, thus an experimental signal in this direction could be
ascribed to the effects of non-standard interaction type described in
this paper. The extension of the contours to small values of the
flavor flux ratios (\eg, $R_{\mu\tau} \sim 0.5$, to be compared with
\fig~\ref{fig:distr}) even in the standard case can be attributed to
the finite resolution assumed for the measurements.

\begin{figure}
  \begin{center}
    \includegraphics[width=0.95\textwidth,clip=true]{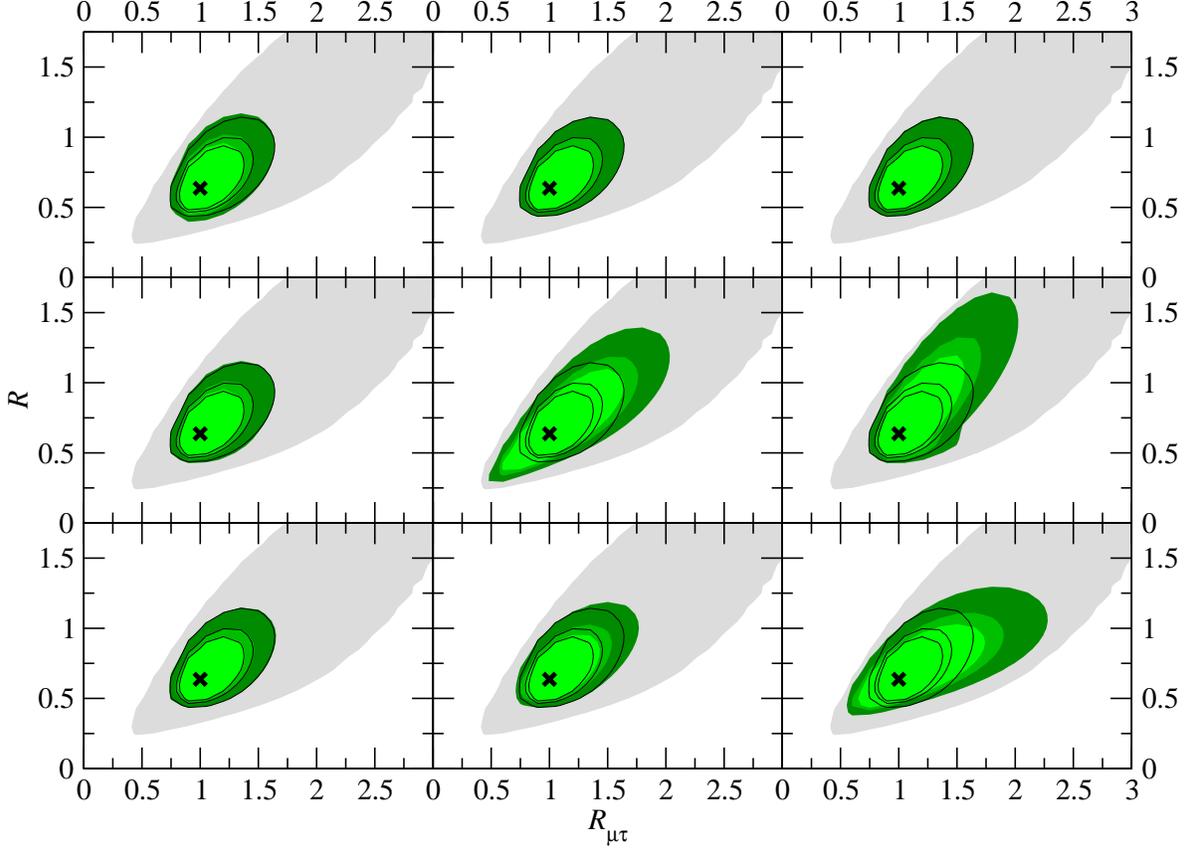}
    \caption{\it Isocontours of $\chi^2_{\rm min}$ in the
      $R_{\mu\tau}$--$R$-plane in the case of a muon-damped pion
      source. The black curves correspond to the case with only
      Standard Model interactions, while the three inner shaded
      regions are produced allowing one $\eps$ to vary with an
      external input error of $0.1$ at $\sig{1}$. The panels are
      placed such that the $\eps$ varied has the same position in the
      $\eps^{ud}$ matrix as the plot has in the figure, \ie, the upper
      right plot corresponds to allowing $\eps^{ud}_{e\tau}$ to vary
      (we put the neutrino flavors in the order
      $\{\nu_e,\nu_\mu,\nu_\tau\}$). The contours correspond to 90~\%,
      95~\%, and 99~\% C.L., respectively. The outer shaded region is
      the 99~\% C.L.\ contour from the full NSI simulation. The black
      crosses correspond to the prediction from tribimaximal mixing in
      the standard framework.}\label{fig:MDPART}
  \end{center}
\end{figure}

It is interesting to understand which of the new parameters that cause
the largest deviations from the Standard Model prediction. This is
illustrated in \fig~\ref{fig:MDPART}, where we show the isocontours of
$\chi^2_{\rm min}$ in the $R_{\mu\tau}$--$R$-plane when allowing for
only one non-zero $\eps$ at a time. From this figure, it is clear that
the most relevant contributions to the extension come from
$\eps^{ud}_{\mu\mu},\eps^{ud}_{\mu\tau}$ and $\eps^{ud}_{\tau\tau}$.
Thus, these are the parameters to which neutrino telescopes would be
most sensitive. Putting stringent bounds on these parameters would
therefore be useful for making more robust predictions for the flavor
flux ratios in the case of the muon-damped pion source.
Note that none of the outer contours from varying individual $\eps$
reach the 99~\% C.L.\ contour from the full simulation. This implies
that in order to accommodate these contours in the full simulation,
some combination of different $\eps$s is needed.

\subsection{Neutron-like sources}

\begin{figure}
  \begin{center}
    \includegraphics[width=0.8\textwidth,clip=true]{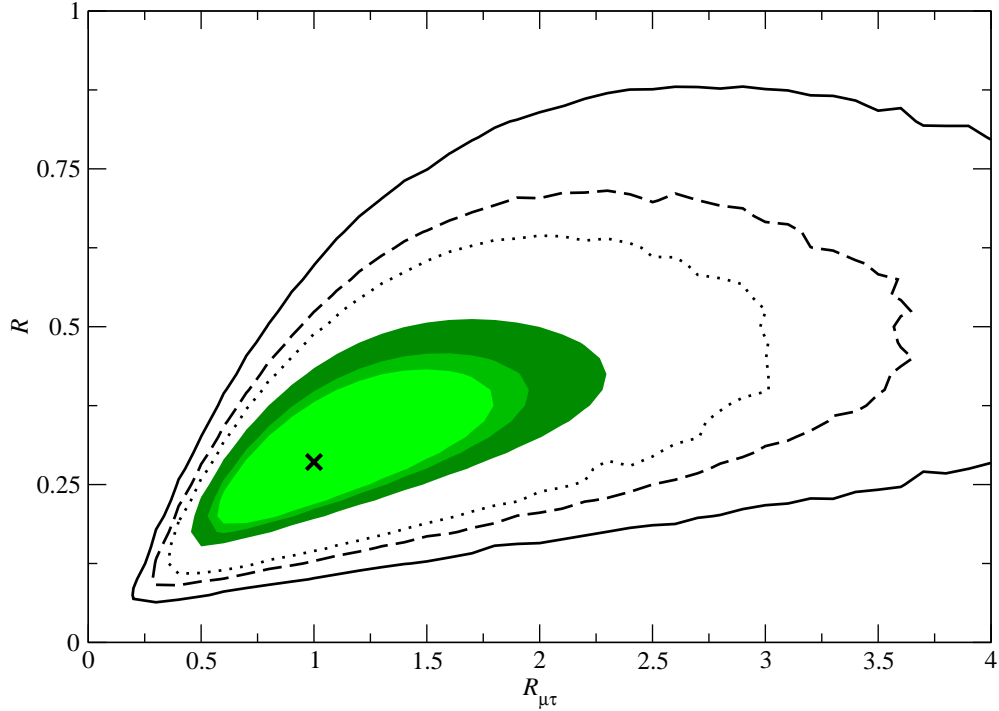}
    \caption{\it Isocontours of $\chi^2_{\rm min}$ in the
      $R_{\mu\tau}$--$R$-plane in the case of a neutron-like source in
      the standard framework for neutrino oscillation and interaction
      (shaded regions) as well as for the framework with included NSI
      (black curves). The contours correspond to 90~\%, 95~\%, and
      99~\% C.L., respectively. The black cross corresponds to the
      prediction in the case of tribimaximal mixing in the standard
      framework.}
    \label{fig:BEFULL}
  \end{center}
\end{figure}

In the case when the source of the astrophysical neutrino flux is
neutron-like (\ie, essentially a beta decay), then the considerations
corresponding to those made for the muon-damped pion sources result
in the flavor flux ratios
\begin{equation}
  R_{e\mu} = \frac{5}{2}, \quad
  R_{\mu\tau} = 1, \quad {\rm and} \quad
  R = \frac{2}{7},
\end{equation}
in the case of tribimaximal mixing. Figure~\ref{fig:BEFULL} shows the
isocontours of $\chi^2_{\rm min}$ in the $R_{\mu\tau}$--$R$-plane for
a neutron-like source using the same procedure as that described in
the previous section. In this case, the extension of the isocontours is
mainly in the direction of increasing $R_{\mu\tau}$, corresponding in
an increase in the $\nu_e$ flux on the expense of $\nu_\tau$ compared
to the standard setup. However, this extension more or less has the
same shape as the extension in the standard case and the big errors
for large values of the flavor flux ratios can be somewhat attributed
to the increasing absolute errors on $\sigma_{R_i}$.

\begin{figure}
  \begin{center}
    \includegraphics[width=0.95\textwidth,clip=true]{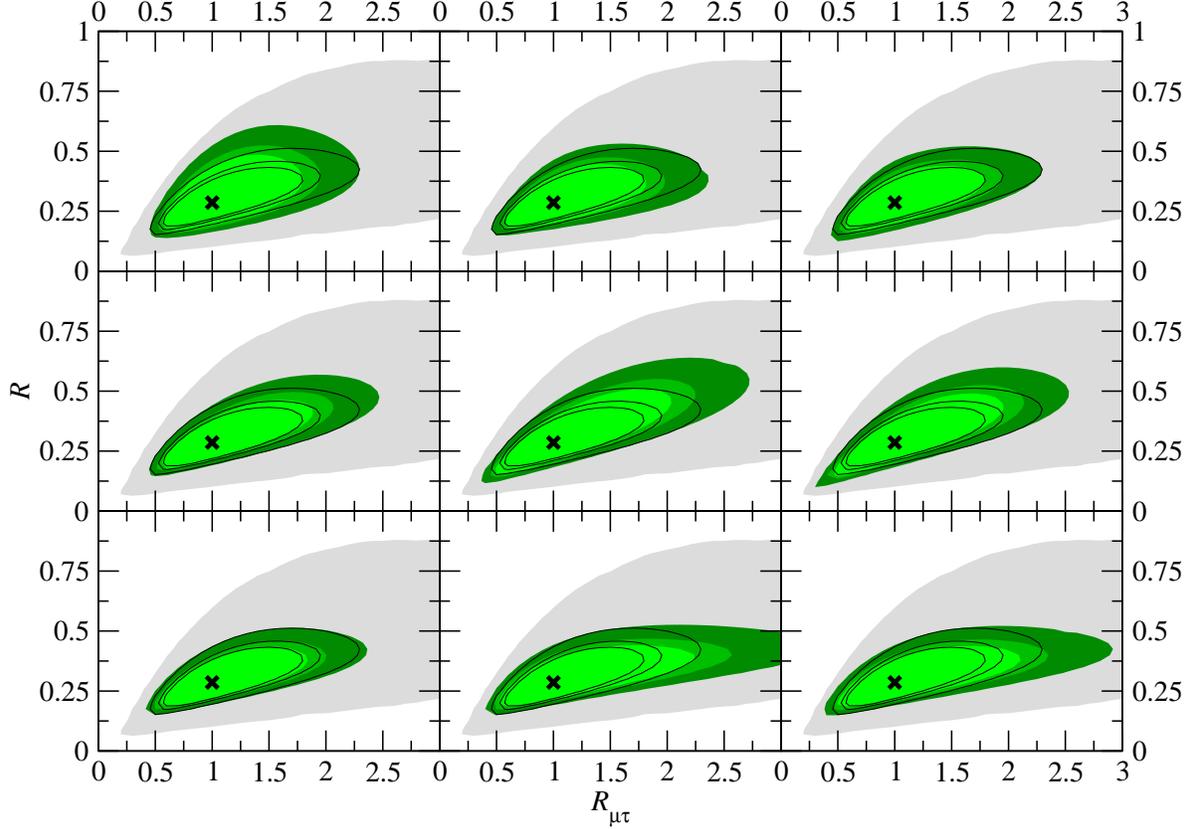}
    \caption{\it Isocontours of $\chi^2_{\rm min}$ in the
      $R_{\mu\tau}$--$R$-plane in the case of a neutron-like
      source. The black curves correspond to the case with only
      Standard Model interactions, while the three inner shaded
      regions are produced allowing one $\eps$ to vary with an
      external input error of $0.1$ at $\sig{1}$. The panels are
      placed such that the $\eps$ varied has the same position in the
      $\eps^{ud}$ matrix as the plot has in the figure, \ie, the upper
      right plot corresponds to allowing $\eps^{ud}_{e\tau}$ to vary
      (we put the neutrino flavors in the order
      $\{\nu_e,\nu_\mu,\nu_\tau\}$). The contours correspond to 90~\%,
      95~\%, and 99~\% C.L., respectively. The outer shaded region is
      the 99~\% C.L.\ contour from the full NSI simulation. The black
      crosses correspond to the prediction from tribimaximal mixing in
      the standard framework.}\label{fig:BEPART}
  \end{center}
\end{figure}

In \fig~\ref{fig:BEPART}, we show the impacts of the individual $\eps$
parameters for a neutron-like source. 
Unlike the case of the muon-damped pion source, with a neutron-like
source we find that all of the $\eps$ has at least some small impact
on the flavor flux ratios, even if it is still relatively small. The
largest difference is observed for the $\eps^{ud}_{\tau\mu}$
parameter, which (together with $\eps^{ud}_{\tau\tau}$) seems to be
the source of the extension to larger $R_{\mu\tau}$ with $R$ being
essentially constant. Also the impact of $\eps^{ud}_{ee}$ is peculiar,
even if it is not as large as that of $\eps^{ud}_{\tau\mu}$. It would
seem that this parameter fixes $R_{\mu\tau}$ while altering $R$,
signifying a change in the flux of $\nu_e$ compared to the other two
species.

\subsection{Non muon-damped pion sources}

As discussed in \Sec~\ref{sec:SDNSI}, when a pion source is not
muon-damped, there will essentially be two different sources of
neutrinos -- the pion decay and the subsequent decay of the
muon. Since the NSI involved in the two different processes do not
depend upon the same $\eps$s, the effective parameter space is
increased quite dramatically, even compared to the previously
considered NSI scenarios. For this type of source, the prediction of
tribimaximal lepton mixing is given by
\begin{equation}
  R_{e\mu} = 1, \quad
  R_{\mu\tau} = 1, \quad {\rm and} \quad
  R = \frac{1}{2},
\end{equation}
simply due to the fact that the flavor fluxes are predicted to be
equal.

\begin{figure}
\begin{center}
\includegraphics[width=0.75\textwidth,clip=true]{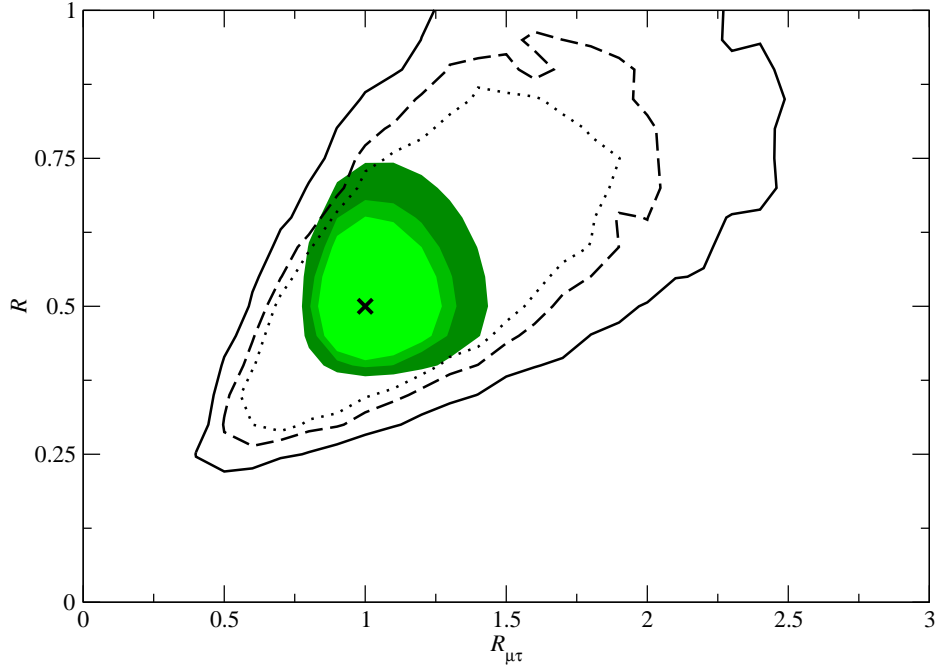}
\caption{\it Isocontours of $\chi^2_{\rm min}$ in the
  $R_{\mu\tau}$--$R$-plane in the case of a non muon-damped pion source in
  the standard framework for neutrino oscillation and interaction
  (shaded regions) as well as for the framework with included NSI
  (black curves). The contours correspond to 90~\%, 95~\%, and 99~\%
  C.L., respectively. The black cross corresponds to the prediction in
  the case of tribimaximal mixing in the standard framework.}
\label{fig:PIFULL}
\end{center}
\end{figure}

In \fig~\ref{fig:PIFULL}, we show the results of the minimization of
$\chi^2$ both for the standard case and when allowing for NSI (both in
the pion and muon decays).
For large values of $R_{\mu\tau}$ and $R$ in the NSI case, this figure
is not very smooth. The reason for this is that the $\chi^2$ is now a
function of a very large number of parameters and the minimization
procedure is not always able to find the actual minimum. In this
scenario, we see that the extension of the contours due to NSI is
mainly along the direction of constant $R/R_{\mu\tau}$, corresponding
to a fixed ratio between the $\nu_e$ and $\nu_\tau$ fluxes while the
relative $\nu_\mu$ flux varies (again, the smaller size of the regions
for smaller $R_i$ is mainly due to the differences in $\sigma_{R_i}$).

\begin{figure}
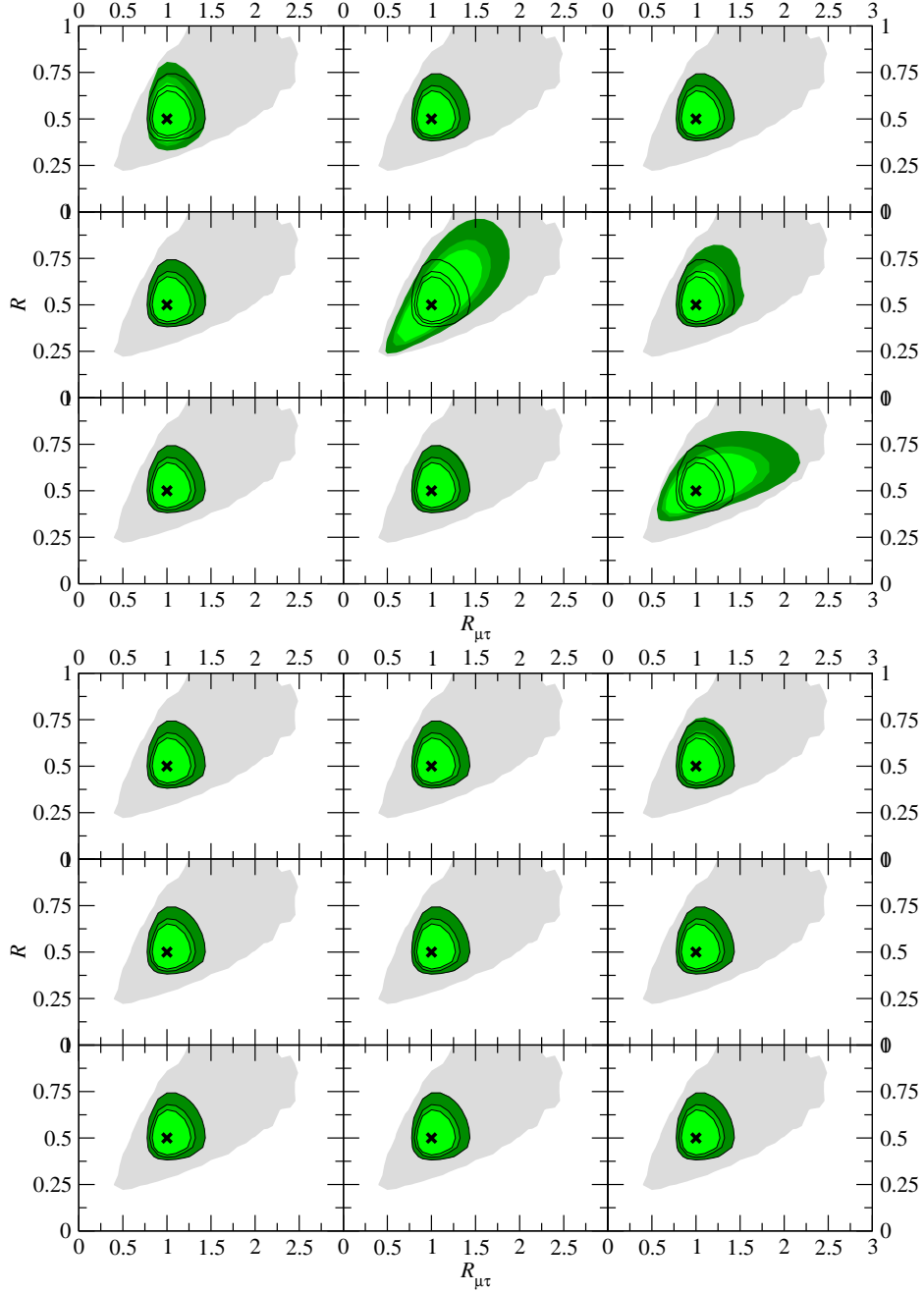

  \begin{center}
    \includegraphics[width=0.75\textwidth,clip=true]{PIPARTUD.eps} \\
    \includegraphics[width=0.75\textwidth,clip=true]{PIPARTME.eps}
    \caption{\it Isocontours of $\chi^2_{\rm min}$ in the
      $R_{\mu\tau}$--$R$-plane in the case of a non muon-damped pion
      source. The black curves correspond to the case with only
      Standard Model interactions, while the three inner shaded
      regions are produced allowing one $\eps$ to vary with an
      external input error of $0.1$ at $\sig{1}$. The upper plot shows
      the dependence on the $\eps^{ud}$ and the lower plot the
      dependence on the $\eps^{\mu e}$. The panels are placed such
      that the $\eps$ varied has the same position in the matrices as
      the plot has in the figure, \ie, the upper right panel of the
      upper plot corresponds to allowing $\eps^{ud}_{e\tau}$ to
      vary. The elements of the figure are the same as in the previous
      plots.}\label{fig:PIPART}
  \end{center}
\end{figure}

As in the previous scenarios, we present the dependence on individual
$\eps$s in \fig~\ref{fig:PIPART}. Here, the exception is that the
parameter space now consists of an additional nine $\eps$s, which we
illustrate by two different plots for the dependencies on the
$\eps^{ud}$ and $\eps^{\mu e}$, respectively.

The effects of the $\eps^{ud}$ parameters are very similar to what
they were in the case of the muon-damped pion source, although a bit
less pronounced. This is to be expected, since the processes of the
muon-damped source also play a role in the case when the muons are not
damped -- although the effect is less pronounced as there are also
other processes involved. However, the effects of the $\eps^{\mu e}$
parameters are essentially negligible, with the only parameter
actually showing any difference (although still a very small one)
being $\eps^{\mu e}_{e\tau}$. Thus, it would seem that the $\eps^{\mu
  e}$ cannot play a significant role when it comes to astrophysical
neutrino fluxes.

\subsection{Comparison of scenarios}

\begin{figure}
  \begin{center}
    \includegraphics[width=0.85\textwidth,clip=true]{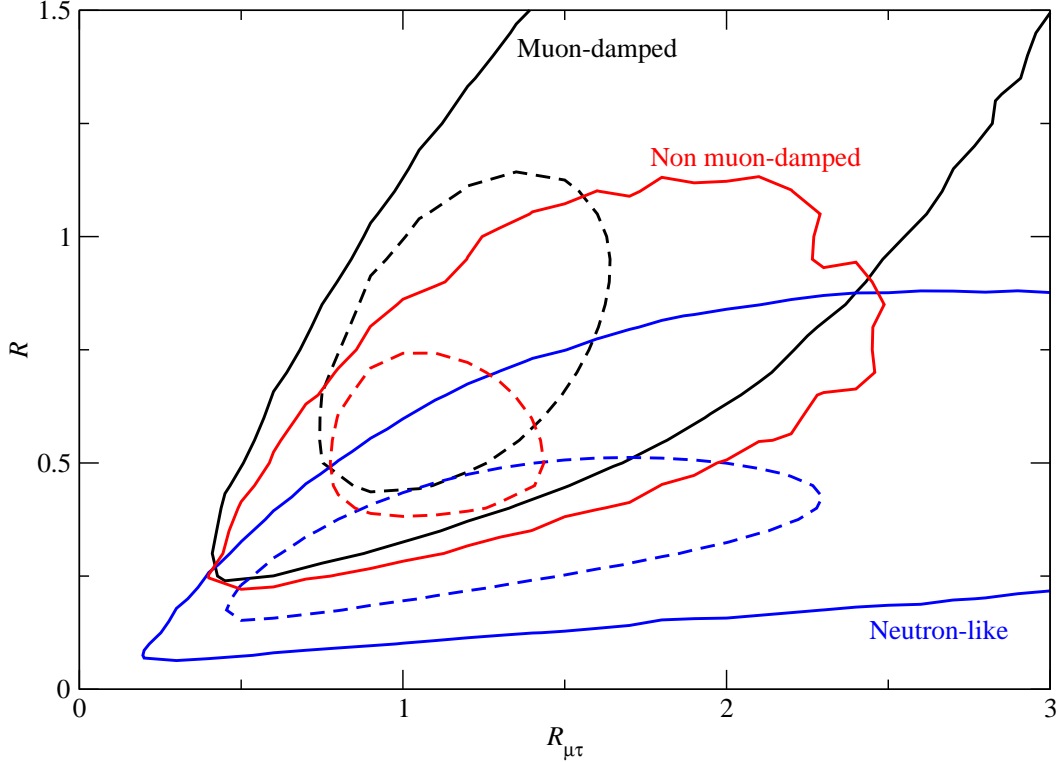}
    \caption{\it The 99~\% C.L.\ isocontours in the
      $R_{\mu\tau}$--$R$-plane for the sources we have discussed. The
      solid lines correspond to the contours when allowing for general
      NSI, while the dashed contours correspond to the result in the
      case of Standard Model interactions only.}
    \label{fig:COMBO}
  \end{center}
\end{figure}

An important aspect in the study of astrophysical neutrino fluxes is
what we could learn about the source from studying the flavor
composition of the flux (\ie, what type of source that is providing
the flux). Thus, in \fig~\ref{fig:COMBO}, we show the 99~\%
C.L.\ contours for all of the source types discussed above.
As can be seen in this figure, the pion sources could be relatively
well separated from the neutron-like sources in the standard scenario,
while it could be harder to tell whether there is muon-damping or not
in a pion source. This conclusion is not significantly altered when
taking NSI into account, although it will be a little harder to tell
the sources apart if flavor ratios in the region expected from
Standard Model interactions only are measured. However, since the NSI
mainly extend the contours in different directions for different
scenarios, large parts of the space of possible measurements allowed
in the case of a neutron-like source are not allowed for pion sources
and vice versa. Still, telling if there is muon-damping in a pion
source remains a difficult endeavour. The contour extension due the
NSI also has another intriguing property. Since the extensions are
mainly not in the direction to where other scenarios are located,
there is a fair chance that if NSI actually appear in the flavor flux
ratios, their effect could not be misinterpreted as simply being due
to a different neutrino source.

\subsection{Impact of experimental uncertainties}
\label{sec:unc}

\begin{figure}
  \begin{center}
    \includegraphics[width=0.95\textwidth,clip=true]{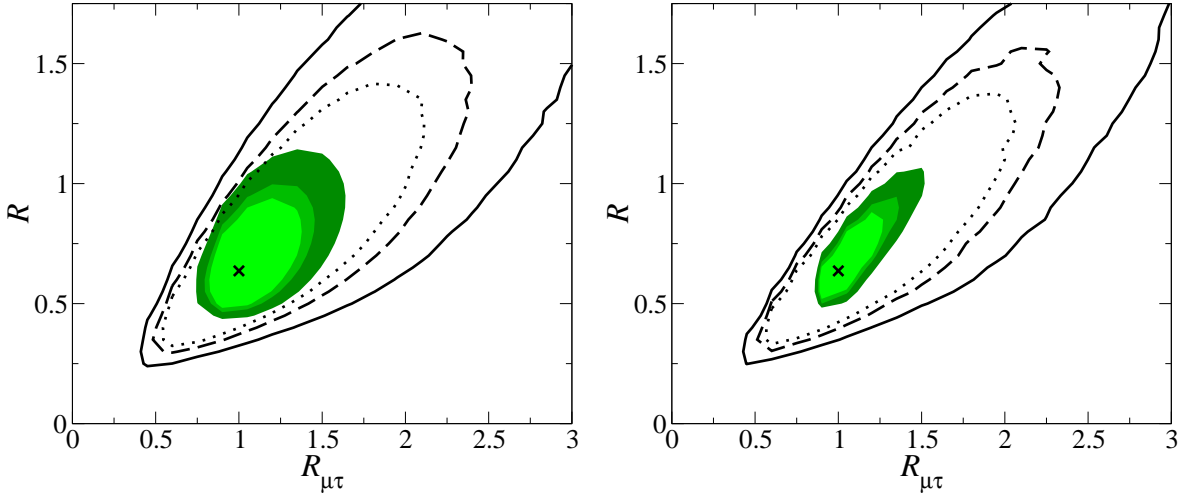}
  \end{center}
  \caption{\it The isocontours in the $R_{\mu\tau}$--$R$-plane in the case
    of a muon-damped source assuming different experimental
    resolutions for the flavor flux ratios. The left panel assumes
    $\sigma_{R_i} = 0.1 R_i^{\rm exp}$ (\ie, identical to \fig~\ref{fig:MDFULL}), while the right panel assumes
    $\sigma_{R_i} = 0.05 R_i^{\rm exp}$. The elements of the plots are the same
    as in \fig~\ref{fig:MDFULL}.}
  \label{fig:MDRES}
\end{figure}

As mentioned earlier, the experimental precision with which the
neutrino flavor flux ratios can be measured is an important part of
our analysis. In our discussions so far, we have assumed a relative
error of 10~\% and it is of importance to understand how this affects
our results. For this purpose, \fig~\ref{fig:MDRES} shows the full
result for the muon-damped source both with our initial assumption of
$\sigma_{R_i} = 0.1 R_i^{\rm exp}$ as well as with a more optimistic assumption
of $\sigma_{R_i} = 0.05 R_i^{\rm exp}$.
As can be seen in this figure, a better experimental resolution makes
the NSI degeneracies more pronounced. This is to be expected, since
better measurements allow for better opportunities to extract
information of the underlying physics. Conversely, if the resolution
is made worse, it will become harder to distinguish the NSI
degeneracies from the standard scenario as deviations could simply be
due to experimental uncertainties.

\section{Summary and conclusions}
\label{sec:S&C}

In this paper we analyzed the impact of new physics effects in the
production and detection of astrophysical neutrinos. Detection
processes were considered independent of the source producing
neutrinos and the possible new physics effects were parametrized in
terms of the matrix $\eps_{\alpha \beta}^{ud}$ which alters the
charged-current reaction $\nu + X \rightarrow Y + \ell_\alpha$
involving leptons and hadrons. The same matrix also intervenes in the
production of neutrinos from muon-damped pion sources (in which
neutrinos coming from subsequent muon decays do not contribute to the
measurable neutrino flux on Earth) and from neutron-like sources. If
muon decays also play a role in the production of neutrinos (as in the
case of the ``standard'' pion sources), then a new set of parameters
$\eps_{\alpha \beta}^{\mu e}$ should be introduced to take the purely
leptonic process into account. Since the fluxes of astrophysical
neutrinos can be measured at neutrino telescopes, we have considered
the experimentally accessible flux ratios
$R_{\mu\tau}=\phi_\mu/\phi_\tau$ and $R=\phi_\mu/(\phi_e+\phi_\tau)$,
in order to investigate the effects of the new parameters. This was
done through a $\chi^2$ analysis to see how well the neutrino flux
ratios at neutrino telescopes can be accomodated in different models,
assuming tri-bimaximal mixing as an external central value for the
standard neutrino oscillation parameters. Making this analysis in the
Standard Model first and then including the NSI gave us insight of how
new physics can affect the possible ranges of these observables. Our
results can be summarized in the simplified scenario, where only one
source at a time is responsible for the neutrino flux on Earth. In
particular, for muon-damped sources, we found that with NSI, the
measured flux ratios can be significantly larger than their
tri-bimaximal mixing values, an effect mainly due to
$\eps^{ud}_{\mu\mu},\eps^{ud}_{\mu\tau}$ and
$\eps^{ud}_{\tau\tau}$. Also for neutron-like sources the extension of
the contour plots obtained including NSI effects are in the direction
of increasing $R_{\mu\tau}$, with $R$ being essentially constant (thus
a significant change in the $\nu_e$ flux is at work). 
Moreover, all the epsilon parameters contribute, in some extent, to
the shift of the flavor flux ratios. The analysis of the non
muon-damped sources is more complicated. This is due to the two
independent classes of new parameters entering in the game. However,
while both $R$ and $R_{\mu\tau}$ can be sizably different from the
Standard Model predictions, we found that the leptonic $\eps_{\alpha
  \beta}^{\mu e}$ have negligible impacts on these results.  In
principle it is possible, in both Standard Model and new physics
scenarios, to learn something about the source, since the
observational parameter space covered by the flux ratios from one
particular source does not overlap significantly with that of another
source.

It should be noted that all of our considerations depend on the
assumed uncertainties on the extraction of the flux ratios as well as
on how well the standard neutrino oscillation parameters have been
determined. In fact, although smaller $\sigma_{R_i}$ makes the NSI
effects more pronounced, larger $\sigma_{R_i}$ will completely
oversahdow any significant new physics effect. In addition, we have
considered relatively large values of the NSI paramters. It should
therefore also be noted that the effect of small NSI will become
increasingly difficult to observe unless the flavor flux ratios and
standard neutrino oscillation paramters can be very accurately
determined.

\section*{Acknowledgments}

The authors would like to thank Tommy Ohlsson and He Zhang for reading
the manuscript and providing useful comments.

This work was supported by the Swedish Research Council
(Vetenskapsr{\aa}det), contract no.~623-2007-8066 [M.B.].
This work has been partly supported by the Italian Ministero dell 'Universit\'a
e della Ricerca Scientifica, under the COFIN program for 2007-08 [D.M.].



\end{document}